# iFUSION: Standards-based SDN Architecture for Carrier Transport Network

L.M. Contreras, Ó. González, V. López, J.P. Fernández-Palacios, J. Folgueira
*Transport & IP Networks - Systems and Network Global Direction*
*Telefónica GCTIO Unit*
Madrid, Spain
{luismiguel.contrerasmurillo,oscar.gonzalezdedios,victor.lopezalvarez,juanpedro.fernandez-palaciosgimenez,jesus.folgueira}@Telefónica.com

*Abstract*—Network softwarization will permit the dynamic configuration of the transport infrastructure, in order to facilitate the adaptation needed for future advanced services. However, the promise of full flexibility can only be achieved through proper levels of abstraction and by means of normalized interfaces and service and device models. This paper presents the SDN architecture that is being defined by Telefónica for the management and control of its transport networks, including a variety of technologies such as IP, optics and microwave radio links, pursuing a standard-based approach for an effective and future-proof introduction of softwarized operations.

## I. INTRODUCTION

Service providers are in a continuous evolution in order to satisfy the changing and variable service demand of the end users. This affects not only to the service offering but also to the network supporting the delivery of such services. A key part of it is the transport infrastructure.

Transport networks are in charge of forwarding aggregated traffic demands from multiple end users consuming different services among different cities, regions or continents. Traditional transport network architectures were conceived and designed having in mind both the characteristics and the traffic demands of the classic services (e.g. Internet access or VPNs), which in the past were considered predictable in terms of origin and destination, as well as their pattern.

Traditional carriers' network operation is considered very complex and not adaptable to flexible traffic requirements. Typically, network operation has been accomplished either directly through manual procedures with direct access to the nodes or leveraging on Network Management Systems (NMSs), when available, as part of the Operation Support Systems (OSSs). Multiple manual configuration actions are needed in distinct network nodes just for a simple capacity upgrade in a path, which prevents network operators of optimizing the resources available, usually forcing to over-dimensioning the network. Noting that a mid-size network can require of few thousands of nodes along its distinct layers (i.e., interconnection, backbone, distribution, aggregation, etc.) this deals to hundreds of thousands of nodes configurations per year. The lack of automation in these actions not only requires from large periods of planning and execution (e.g., scheduling of maintenance windows, careful elaboration of configuration templates, etc.) but also are prune to errors, slowing down the pace of deployment of new services and capacity upgrades.

Alternatively, when available, NMSs have assisted in these operational tasks facilitating the work especially at the service provision time for the configuration of protocols and parameters, collection of alarms, etc. However, network equipment solutions from different vendors typically use vendor-specific NMS implementations, then not permitting a single pane of glass for network operation except through the integration of very complex umbrella systems, with specific and often proprietary interfaces, which require constant integration efforts along the lifetime of the equipment (ant is accompanying NMS).

These same alternatives has been usually replicated in each technology silo in a network operator. For instance, separated organizations have been in charge of the planning and engineering of IP, optics and wireless transport technologies in a network, acting as interrelated but isolated environments, each of them with the same mode of operation. That operational separation reproduces the same concerns at each technological silo. Thus, for instance, it can derive in complex and long workflows for network capacity provisioning (e.g. up to two weeks for Internet service provisioning, and more than six weeks for activating core routers connectivity services over a photonic mesh).

This mode of operation is no longer valid nor sustainable, especially with the generalization in the support of cloud and virtualization technologies, as advanced in [1]. This will become even more evident once the new generation of mobile communications, 5G, start to be deployed, with new paradigms such as network slicing for adapting to specific service characteristics.

Network operations are heavily influenced by the control and management capabilities available. Key aspects of operations such as network provisioning or troubleshooting can then benefit from advanced tools and mechanisms. It is then necessary to perform a transformation process in both the network and the OSS landscape. The network transformation should provide a simplification in the architecture by reducing the number of layers and consolidating network functions in an smaller number of nodes, then reducing the cost of network upgrade and reducing the number of nodes to be operated in the network. The OSS transformation should provide a higher level of automation with a unified number of components such as inventory, activation, performance and fault management tools. In order to fully unlock the potential of such transformations, the paradigm of Software Defined Networking (SDN) becomes the central point of network evolution, covering the majority of existing gaps.

This journey has been initiated by a number of major operators in the world [2][3]. However, the key aspect for a cost-efficient transition to the future mode of operation enabled by SDN is yet open. This is the clear definition of open, standardized interfaces and models that could permit an straightforward and seamless integration of nodes and systems. This paper presents the Telefónica approach towards and standards-based architecture for softwarized network operation. Section II introduces the motivation for this transition as well as the use cases enabled through SDN. Section III described the open SDN architecture proposed by Telefónica, branded as iFUSION, including the reference standards considered. Section IV discusses the main challenges being faced for the real introduction of this architecture in the Telefónica's operations in the world. Finally, Section V ends the paper with some concluding remarks, directions taken and next steps.

## II. MOTIVATION AND USE CASES

Software Defined Networking (SDN) originally [4] entailed the decoupling of control and forwarding planes in the network elements, with the deployment of a centralized controller with the complete network view, running intelligent algorithms and applications (either as part of the controller or on top of it) and instructing the nodes accordingly. This original view, when ported to real carrier networks, has been evolved slightly towards an architecture where the centralized control assists the network elements on the forwarding tasks, providing single and unified control environment for network operations and at the same time optimizing the usage of network assets. The network elements yet retain control capabilities (in some cases alleviating some signaling tasks) but leveraging on the centralized controller for end-to-end and cross-layer actions through programmable interfaces.

The use of the SDN principles is aimed to simplify the network operation and allowing a fast reaction and adaptability to network changes motivated by traffic variability or simply because of service configuration. Besides network element control functions, SDN is being considered also as a mean to provide support for management functions, such as collection of real-time information that could permit the automatic configuration creation and activation in network elements, as triggered by the OSSs. This section introduces the main motivation for this transition as well as details a number of use cases enabled by this new architecture.

### A. Motivation

The operation of a network offers multiple challenges due to the heterogeneous nature of the equipment constituting it, in terms of architecture, technology and implementation (i.e., different manufacturers develop similar but not uniform solutions). Among those challenges, it may be mentioned:

- Complex and intricate procedures for service delivery: dependencies among services in the network cannot be detected in advance, nor optimal usage of resources can be anticipated.

- High customization in the configuration during service creation. Even for a single transport technology, the specificities in the design of each per-vendor implementation (in terms for instance of command structure, configuration options or parameter values) force a constant customization of the services constructs. This affects not only the provision phase, but also the operation and maintenance of the services.

- Slow adaptation of the network to changing demands. Since fast reaction to network conditions is not feasible with existing systems, capacity overprovision remains as the unique way of avoiding service affection. Such overprovisioning is intensified even more when considering the forthcoming unpredictability on traffic patterns due to the flexibility granted by the virtual computing capabilities, for hosting any kind of traffic source.

- Long time-to-market. Overall, the competitiveness of the network operator is impacted since market and commercial opportunities do not have enough agility and flexibility to adapt to the customers' demands. SDN will enable that network services are implemented by programming the network instead of re-architecting it, as often happens nowadays.

In addition to that, in the particular case of Telefónica, group formed by a number of distinct domestic networks in more of a dozen of countries, there exists the additional difficulty on defining and standardizing services across the group. This lack of uniformity forces to customize the implementation of a service for each different operation.

These facts together with a number of new use cases, as described below, foster the need of evolving the transport network towards a full programmable and automated environment.

### B. Vendor-agnostic operation

Network creation and service activation are performed through configuration of network elements. However, such configuration is vendor-dependent because of disparity on implementations from distinct manufacturers. In consequence, when a service needs to be configured through different networks, the configuration process need to be done in each sub-set of network elements separately.

In addition to that, the integrating new equipment from a new vendor is time-consuming, needing changes in the OSS tools already deployed. Such longer processes delay the introduction of new technologies, de facto slowing down the transformation process and the agility needed by operators.

One of the benefits of including a SDN solution is to speed-up the processes of integrating new vendors (or new OSS systems) in the network. To do so, it is required a standard NBI towards the OSS systems (network planning tool, inventory database, configuration tool, etc.) and a standard SBI towards the network elements that could depend only on the network technology/segment (e.g. microwave, Metro-IP or optical) and actually not in the vendor's implementation.

Another advantage of having a SDN Solution deployed for this process, is that the Inventory System can be better fully synchronized with the network so the provisioning can be done based in the status of the network, avoiding any misalignment between the planning process and the deployment process.

Vendor-agnostic operation is fundamental to have a common way of controlling and managing the transport infrastructures. SDN, through the deployment of fully standard South-Bound Interfaces (SBI) towards the network elements performing forwarding, and North-Bound Interfaces (NBI) to OSSs and other management elements, and it is capability to abstract network resources to upper layers, represents an important enabler.

*C. Traffic Engineering and Path Computation*

Traffic engineering (TE) allows the enforcement of traffic steering flows by leveraging onto MPLS tunnels or Segment Routing paths. This permits to increase the efficiency on the use of the network resources by properly mapping the traffic flows to the available resources, and improve network management, including troubleshooting, to overcome difficult failure situations. Increasingly complex network scenarios such as large single domain environments, multi-domain or multi-layer networks require the usage of algorithms for efficiently computing end-to-end paths.

This complexity is driving the need for a dedicated SDN controller, which will perform path computations and be adaptive to network changes. The Path Computation Element (PCE) function allows performing complex constrained based path calculation over a network graph representation. The centralized path computations introduced by the PCE, improves the application of TE policies in MPLS and GMPLS networks by mitigating race conditions inherent of distributed systems.

*D. End-to-end network automation*

When network operators consider to deploy new network services, it is common that the service is implemented on top of mixed with the network elements configuration. Therefore, this ends up in a scenario where even overlay services are very difficult to deploy. It seems convenient to have a clear definition of services and network, so the operator can deploy services across multiple technologies and even administrative domains though a common network services API.

The services can be decoupled from the specificities of the underlying transport capabilities through abstraction. However, it is necessary to ensure that proper mapping exists between service requirements and transport capabilities. Once decoupled, coordination between service-related and transport-related SDN control functions is required, while de-coupled, facilitating differentiated evolution of both, as defined in [5]. Modular control approach for both services and transport/connectivity is easier to design and validate, considering cooperative interaction between both levels. Control solutions on either side can evolve independently, not only creating less risk but allowing for independent and optimized migration.

III. OPEN SDN ARCHITECTURE FOR TRANSPORT NETWORKS

End-to-end provision and operation of the network requires a network-service interface, being a common standard interface for an easy integration of elements from distinct manufacturers, supporting the same semantic. The present mode of operation does not have a common interface to support deploying multiple services. Additionally, OSSs and NMSs have multiple vendor-specific interfaces, which creates great problems in terms of tools integration.

With the current approach, it is not easy to provide an abstracted topology view or service-specific view of the network to the application in a generic fashion, or to allow application to request and/or control virtual network resources. For an efficient and consistent programmability of network devices, network resources should be abstracted towards upper control layers, so that any unnecessary complexity of any particular implementation is hidden and only dealt with in the appropriate layer.

An open SDN architecture for transport networks should use standard network configuration interfaces, allowing to trigger automated standard control plane for multi-domain, multi-vendor or multi-layer operation. Key building blocks of such open, unified network provisioning architecture are:

- Interface towards network elements: multi-vendor nodes configuration should be accomplished by standard interfaces.

- Service layer and network coordination: coordinated network and service layer according to service requirements (e.g. service requirements depends on applications).

- Support of abstraction in a Network-Service interface: definition of APIs hiding no relevant details of the network. This enables having a common entry point to provision multiple services.

With this principles in mind, Telefónica has defined an open SDN architecture for transport networks, named iFUSION, and described next.

*A. Global architecture*

iFUSION is a reference model architecture permeating the separation of concerns for both network and service layers. The proposed architecture can be shown in terms of components and relationship among them, as depicted in Figure 1.

iFUSION proposes the use of standard North bound interfaces leveraging on RESTconf/YANG and standard data models based on latest developments in IETF, ONF and OpenConfig organizations.

The key elements of the SDN iFUSION architecture are the following:

- <u>SDN Domain</u>: It is a set of network elements under the supervision of the same SDN Controller. There are several possible levels in the decoupling of control and data planes. The preferred level of decoupling in Telefónica depends on the network technology. For example, in the case of MPLS, the network element runs the distributed protocols (e.g. IS-IS TE, RSVP-TE) and the controller only needs to configure it.

- <u>SDN Transport</u>: It is the whole network controlled by following SDN principles. It is divided into SDN Domains for technology/scalability/administrative principles. A SDN Transport Controller (also referred as SDN Orchestrator), will take care of stitching the different domains/layers/technologies.

- <u>SDN Domain Controller</u>: This element is in charge of a set of network elements. It has standard southbound interfaces that depend on the technology, but not in the equipment vendor, to communicate with the network elements. It also has a northbound interface to communicate with the SDN Orchestrator and the OSS.

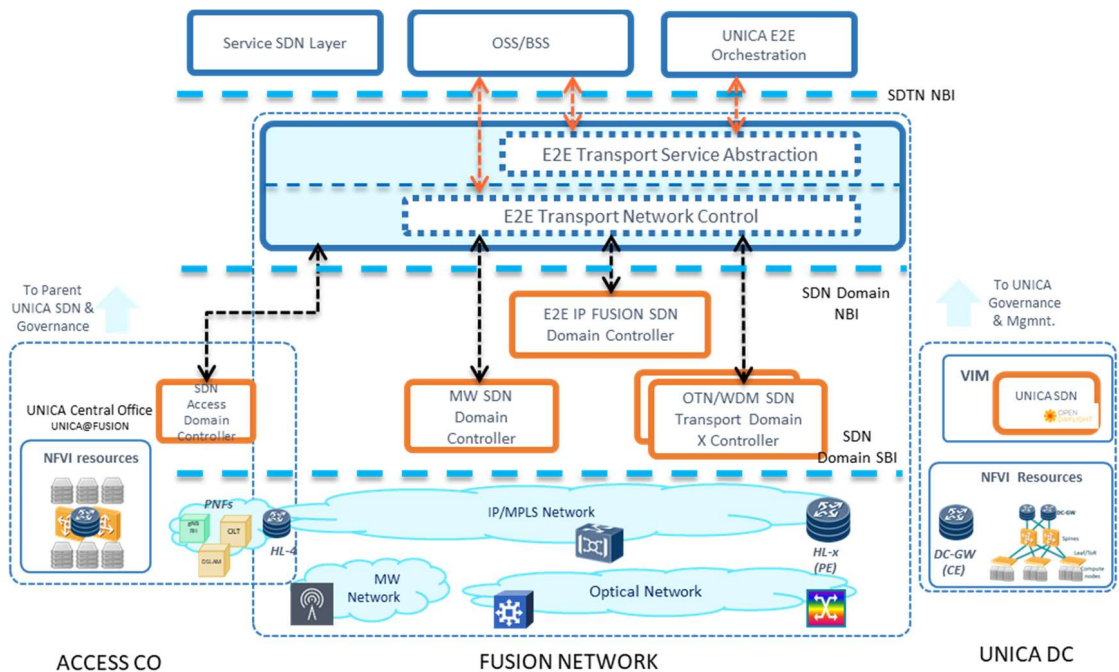

Figure 1. iFUSION architecture

- Software Defined Transport Network (SDTN) Controller: In case several SDN Domains are needed, the SDN Transport Controller is in charge of providing services through several domains.
- Southbound Interface: It is the interface, based on a standard, between the SDN Domain Controller and the Network Element. Not only the communication protocol needs to be standard, but also the data model used.
- Northbound Interface: If is the interface, based on a standard, between the SDN Domain Controller and the OSSs and SDN Transport.
- Service SDN controller: An additional SDN layer that takes into account services might be needed.

As of today, one single SDN implementation does not fit the entire network. An End-to-End SDN Controller for the whole Network layer is not perceived as feasible due to technological and operational issues. Short and medium term scenarios will be conformed by a set of Domain Controllers, each one per network segment (SDN Domain). The iFUSION architecture is designed as a hierarchical model where each network segment is controlled by a dedicated SDN Domain controller. The transport network, due to its wide scope and complexity, is divided in three main technology domains: IP, Microwave (MW) for wireless transport, and Optical for transmission.

The Software Defined Transport Network (SDTN) Controller is responsible to orchestrate the respective SDN Domain controllers within the transport segment (IP, Optical and MW) through the Domain Controllers' NBI, providing an end-to-end transport network vision. The SDTN Controller aggregates demands from the management and services layer exposing a unified NBI which should provide resource configuration abstraction and technology agnostic service definition. The SDTN entails two main building blocks: *(i)* end-to-end Transport Network Control and *(ii)* end-to-end Transport Service Abstraction.

The SDN Domain controllers, on the other hand, are in charge of all the devices in the domain. Each SDN Domain controller unifies the device configuration interface and provides vendor-agnostic network configuration, monitoring and resource discovery. Besides, the Domain Controller exposes high-level network services abstraction to OSS and BSS layers through its North Bound Interface (NBI). Therefore, the abstraction of device specific configuration from network service definition is one of the main features that the SDN controller implements. Moreover, the SDN Domain Controllers entail the function of Path Computation Element to manage and optimize traffic engineering in the domain.

The specific SDN requirements and implementations vary depending on the network environment it is applied to:

- Different network equipment to control (IP/MPLS routers, DC switches, OpenFlow switches, OTN nodes, ROADMs, Microwave devices, CPEs…)
- Different technical realizations (physical elements vs. virtual elements)
- Different services to configure and control (VPNs, tunnels, optical paths, Layer 2 services, Layer3 services, overlay service tunnels, …)

SDN technology is also the base of internal connectivity inside the virtual infrastructure domains, as the case of UNICA, the virtualization infrastructure for Telefónica [6]. Some mechanisms to coordinate with the Transport SDN will be required. The interaction will be done at a horizontal level between the WIM (Wide area Network Infrastructure Manager) [7] and the Transport Controller.

The following sections described in detail the approach followed per technological domain.

## B. IP domain

IP networks are deployed following a hierarchical model, mixing equipment from different vendors. The IP boxes are interoperable at data plane level and control plane level (e.g. routing protocols such as IS-IS, OSPF or BGP). Due to scalability reasons, the IP networks are typically subdivided in IP domains, so the routing and control protocols are confined to their respective domains.

The foreseen SDN solution for IP segment is based on a single, multi-vendor IP SDN Domain Controller in charge of configuring the IP network elements, as shown in Fig. 2.

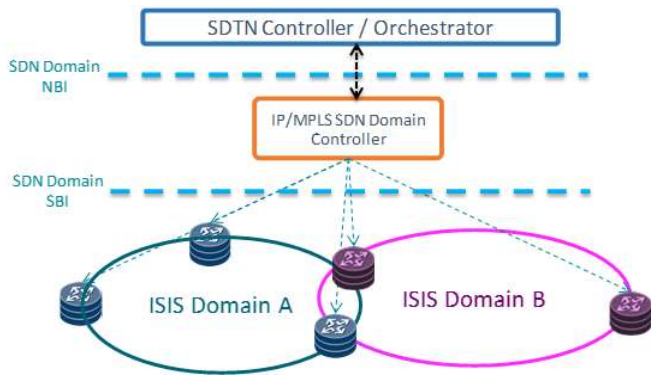

Fig. 2. Single multi-vendor IP SDN Domain Controller deployment

The target SBI for vendor agnostic device configuration shall be compliant with NETCONF [8] standard protocol. The set of device configuration data models (SBI) for the IP segment is still under definition but shall be defined in YANG [9] modelling language, while it could be inherit or complemented from/by other alternative modelling tools, such UML, to help its definition.

From the implementation perspective, it is assumed that devices might not natively support the data models selected, and thus a mediation layer could be required to be implemented between the controller and the devices. Such a mediation layer would be considered as an interim solution and, if present, shall be transparent without impacting on the performance and capabilities of the defined interface.

Additionally to pure device configuration, the IP SDN Domain controller shall perform Traffic Engineering and Path Computation. With that purpose, some standard and mature control protocols such PCEP and BGP-LS for MPLS networks, shall be implemented to complete the definition of the SBI. As a result, Telefónica expects that the IP SDN controller will assume the control/management of:

- Device configuration of interfaces (VLANs) and routing protocols (BGP, ISIS…)
- Traffic Engineering of MPLS tunnels (LSPs).
- Overlay networks services (L2/L3 VPNs) device configuration (VRFs,...)

The IP SDN Domain controller will be the main entry point to the network elements, to avoid overloading the elements and providing a coherent view. The NBI of the controller will also be based on standard models defined in YANG and implemented either on NETCONF or on its lightweight version RESTCONF [10] with XML/JSON encoding. The NBI shall provide to higher entities within the SDN hierarchy:

- Device inventory information.
- A layered topology view (L2/L3, MPLS) of its controlled network entities.
- LSPs provisioning and path computation.
- Device abstraction for network services towards the SDTN Controller, i.e., for overlay services VPNs (L2, L3)
- Network state and performance monitoring information of the IP domain.

Apart from the previous network segments, SDN will also be introduced in other network environments related with Virtualization and the evolution of the Central Offices [11].

## C. Optics domain

Transport WDM networks from different system vendors are deployed on a regional basis, either for technology redundancy, due to different optical performance requirements (metro vs. long-haul), or simply for commercial reasons.

Without line-side interoperability of the different WDM transceivers and Reconfigurable Optical Add-Drop Multiplexers (ROADMs), there is not a competitive advantage on a uniform configuration interface of the optical devices, since they cannot neither be mixed in a multi-vendor scenario, due to the fact that both line systems and transceivers must be from the same vendor.

With this in mind, in the short term, Optical SDN controllers are expected to provide network programmability and interoperability towards upper layers (multi-layer) and between vendors (multi-domain, multi-vendor) through the support of standard NBIs (i.e. coordination will be provided by upper layer hierarchical SDN controller) [12]. This short term approach will enable the setup and tear down of connections in optical channels (OCh and ODU layers), the discovery the network resources to compose a layered uniform view based on the OTN hierarchy, and the monitoring of the optical network.

The SDN architecture proposed is compatible with a legacy control scenario where a distributed GMPLS control plane has been already deployed. GMPLS control plane can be centrally managed by a SDN domain controller by well-know and mature control protocols, such as PCEP, OSPF and/or BGP-LS already supported in GMPLS devices, beneficing the gradual introduction of SDN. However, current NMS solutions shall evolve to, or co-exist with, the SDN Controller model, enabling network programmability through its NBIs while keeping the current offered features for network creation, resources discovery and monitoring and service creation for L0/L1 layers. Standardization efforts targeting the definition of standard NBIs that can facilitate multi-vendor interoperability (by maintaining administrative domains for each vendor) such as ONF Transport API (T-API) [13] and IETF models [14] are the more promising definitions to implement such capabilities by abstracting the specific configuration of current distributed control planes embedded in Automatic Switched Optical Network (ASON) architectures.

iFUSION relays on ONF Transport API 2.0 as as the reference NBI for the SDN implementation in the optical transport segment, having been experimented in several proof of concepts [15].

In the medium and long term, the direct programmability of the components can have interest in Point-To-Point, Metro and Regional scenarios, where disaggregation of optical transceivers and line side components can play an important role. In this line, OpenROADM [16] and OpenConfig [17] projects have already defined device configuration models for transponders and open line systems. Telefónica is approaching this transformation of the optical control in two phases:

1. Partial disaggregation, as a medium term objective, where the target is to define a standard interface based on NETCONF/YANG, which allows of the Optical SDN Controller to manage third-party terminal devices (i.e., transponders) that can transmit over the vendor line system.

2. Full disaggregation, in the long term, where the objective is the open management of the line system, i.e., the defragmentation of the optical transport network in vendor islands by the adoption of a common standardized interface for open line systems (multi-vendor) to be managed by a single optical SDN Controller.

### D. Wireless transport domain

Wireless Transport networks, typically consisting on microwave (MW) radio links, are deployed on a point-to-point basis covering a given region using several vendors. Currently the wireless networks are operated through vendor proprietary Network Management Systems (NMS) with specific proprietary interfaces.

Operation, configuration and maintenance activities are performed manually and statically. Furthermore, this diversity in NMSs and vendor installed base prevents from using advanced applications that could provide more sophisticated features (e.g. power management or multi-layer coordination [18]), since actual integration costs disincentive any effort in such direction.

This reality has fostered the standardization of a common framework [19] for definition of a unified and standard control plane for microwave systems, pursuing as objective the multi-vendor interworking, multi-layer control, and network-wide coordination.

For practical SDN deployments in this technical domain the ONF has released a standard model published as TR-532 document [20], with the support of the wide community of MW system manufacturers. A number of PoCs have served as validation steps of the viability of the model, as well as helping to refine it and to prepare the industry for this evolution. This is the data model adopted by Telefónica in the iFUSION architecture.

The proposed model is aligned with the view of the SDN Domain Controller as a vendor-agnostic configurator of the MW network. It leaves the service definition to upper applications which need to provide the intelligence to deploy and manage end-to-end services, by configuring every device involved individually through the SDN controller.

Interestingly, one of the outcomes of these PoC has been the definition of a Mediator, in principle devoted to experiment with the model, but later on identified as a mean of integrating legacy MW systems with an evolved SDN-based deployment, then protecting investments and ensuring smooth transition towards SDN.

### E. Integration of SDTN in the overall operator's systems architecture

The SDTN Controller will keep visibility of all the transport network segments. It will expose an abstracted topology view of the network resources and the available set of network services to different clients through its North-Bound APIs.

One of the main drivers of deploying an SDTN controller is service automation. SDTN will enable it, progressively, facilitating that services and network configurations carried out manually today become automated and available through this abstraction layer. The level of abstraction can be different according to the needs of the northbound client (e.g. OSS, service orchestrators/SDN controllers, NFV orchestrator, etc.).

The information exported through the NBI towards OSS and other platforms will cover progressively a number of functional areas. The service's provisioning within the Resource Lifecycle Management (RLM) domain will be the first set of functionalities adopted by the SDTN controller, which will progressively include Performance Management (PM) and Network Planning and Design (NPD), and finally Fault Management (FM) and Resource Inventory Management (RIM) areas which includes the major vendor specific management information will be included in the SDTN. The inclusion of these functional blocks is conditioned to the standardization of the required data models for the SDTN NBI and SDN Domain controllers SBI.

On the SBI of the SDTN, each technology Transport Domain SDN Controller shall expose vendor agnostic network level programmability and resource discovery functionalities. The SDTN Controller SBI is intended, but not limited, to provide access to device's configuration data, to expose per-OSI layer topology and network inventory information, and to offer active monitoring of device configuration changes and network state data (i.e., traffic statistics). Alarm and device inventory information for FM and RIM respectively, is intended to be managed at the SDN Domain controllers in a first phase, but its exposure through the SDTN will be evaluated too.

## IV. CHALLENGES FOR THE DEPLOYMENT OF iFUSION APPROACH

The iFUSION architecture described before, based on standard interfaces and models, represent a future proof approach in the migration towards softwarized networks. It is an efficient approach since going to standards solutions can reduce the integration costs and timing, also permitting, in principle openness by allowing the participation of more industrial players on the final implementation.

Being this a sustainable approach, if faces however a number of important challenges, namely:

- Adoption of the models by the industry: despite the standardization of service and device models, the grade of adoption varies in the industry, in some cases limiting the real alternatives in terms of available implementations in the market. A second risk in the development of the models by the vendors is the fact that implementations of the models could differ in functionality such as presenting distinct behavior for a given actions (e.g., different responses to configuration errors). Finally, usually the models can be augmented with proprietary extensions which can difficult in the future a uniform manner of configuration when multiple vendors are present in the network.

- Model diversity and fragmentation of standardization fora: an extremely rich variety of models is emerging in different standardization bodies, some of them addressing the same target technology. This fragmentation can impact the industry that can diverge in the path towards softwarization, which clearly is a negative effect, since the dominance of one approach over the others can take long time, slowing down the advent of programmability. While in some cases this can be consequence of commercial strategies, in some others is just because the existence of overlapping fora.

- Gaps in controller's NBIs: the transition towards programmability started by exploring the device programmability, then focusing on controller's SBIs. Since the standardization of functionalities and features of network equipment has been reasonably well defined, such approach has progressed in time. However, because of the existence of multiple proprietary and open source controllers, it has not been so easy to identify common functions nor features to be supported by the controllers. In consequence there is not a common approach on NBIs, making difficult the generalization of applications or tools running on top of them for assisting in the network operation and optimization.

## V. CONCLUDING REMARKS

There are a number of problems with the current transport network provisioning approach. First, the interfaces between the service management systems and the umbrella provisioning system are typically proprietary, non-programmable and closed interfaces that prevent new applications from a rapid and automated introduction. Second, the orchestration capabilities across different NMSs (e.g., IP/MPLS NMS and Optical Transport NMS) are very difficult to achieve, as each NMS is a highly specialized vendor element that lacks interoperability with other vendors' elements especially on the NMS to NMS communication. Third, there has been little standardization on interface for upper layer applications or services.

Telefónica is currently working on the full definition of Software Defined Transport Network model, named iFUSION, as main pillar of the network transformations, which covers the SDN-based control for core, metro and backhaul network segments, including IP/MPLS, optics and microwave technologies. For doing so, the approach taken has been the one of relaying on standardized interfaces and models, choosing the more matures in the industry for an efficient introduction of the network softwarization principles, starting 2019.

However, there are yet a number of challenges to solve that require from more collaboration in the industry for the definition of the aforementioned models and standards.

This paper describes the directions taken by Telefonica with regards the three main technologies present in a transport network, that is, IP, optics and microwave radio.

Future work is focused on the deployment of the architecture here described and in the contribution to the selected standardization fora for the complete definition of a standards-based SDN architecture for carrier transport network.

ACKNOWLEDGMENT

This work has been partially funded by the EU H2020 5G-Transformer Project (grant no. 761536).